\documentclass[11pt,a4paper]{article}
\pdfoutput=1

\usepackage{amsmath}
\usepackage{amsfonts}
\usepackage{amsthm}
\usepackage{amssymb}
\usepackage{amscd}
\usepackage[british]{babel}
\usepackage{graphicx}
\usepackage{caption}
\usepackage{subcaption}
\usepackage{psfrag}
\usepackage{epsfig}
\usepackage{rotating}
\usepackage{times}
\usepackage{color}
\usepackage{ulem}
\usepackage{color}

\theoremstyle{plain}

\newtheorem{remark}{Remark}[section]

\newcommand{\boxend}{\flushright{$\Box$}}

\renewcommand{\tilde}{\widetilde}

\begin{document}

\title{Bouncing cosmologies in geometries with positively curved spatial sections}

\author{
Jaume Haro$^{a,}$\footnote{E-mail: jaime.haro@upc.edu} }

\maketitle

\begin{center}
{\small
$^a$Departament de Matem\`atica Aplicada I, Universitat
Polit\`ecnica de Catalunya \\ Diagonal 647, 08028 Barcelona, Spain \\

}
\end{center}

\thispagestyle{empty}


\begin{abstract}
Background boucing cosmologies, driven by a single scalar field,  having a quasi-matter domination period
during the contracting phase, i.e., depicting  the so-called Matter Bounce Scenario,  are reconstructed for cosmologies with spatial
positive curvature.
These cosmologies lead to a nearly flat power spectrum of the fluctuation curvature in co-moving coordinates for modes that leave the Hubble radius during this quasi-matter
domination period, and whose spectral index and its running, which are related with the effective Equation of State parameter given by the quotient of the pressure over the energy
density, are compatible with experimental data.
\end{abstract}

\vspace{0.5cm}

{\bf Pacs numbers:} 04.20.-q, 98.80.Jk, 98.80.Bp

\vspace{0.5cm}










\vspace{0.5cm}


\vspace{0.5cm}


\section{Introduction}
Bouncing cosmologies  (see \cite{brandenberger} for a review) do not have the horizon
that appears in Big Bang cosmology  \cite{aho} and, when the bounce is symmetric,  improve the flatness problem
(where spatial flatness is an unstable fixed point and fine tuning of initial conditions
is required), because the contribution of the spatial curvature decreases in the contracting
phase at the same rate as it increases in the expanding one (see for instance
\cite{brandenberger1}).
Therefore they could, in
principle, be a
viable alternative
to the inflationary
paradigm \cite{bran}. On the other hand,
it is well known that when the background is the Friedmann-Lema{\^\i}tre-Robertson-Walker (FLRW) geometry and one has a single scalar field filling the Universe, within General
Relativity (GR),
only geometries with positive spatial curvature could lead to bounces. However, the most usual way to obtain bounces is to work in the flat FLRW space-time  and to
introduce
nonconventional matter fields
\cite{gost}
in order to break down
 the weak energy condition $\rho+P>0$ (being $\rho$ the energy density  and $P$ the pressure), or to go beyond GR and to deal with theories such as Loop Quantum Cosmogy (LQC),
where holonomy corrections introduce a quadratic correction in the Friedmann equation leading to a Big Bounce  that replaces the Big Bang singularity (see, for instance, \cite{LQC}),
modified $F(R)$
gravity \cite{odintsov} or teleparallel $F(T)$ theories \cite{aho,haroamoros}.

Once one has a bouncing background, the next step is to deal with cosmological perturbations. There is a well known duality between a matter domination epoch in the contracting phase and the
de Sitter regime in the expanding one \cite{wands}, thus a quasi-matter dominated Universe when modes leave the Hubble radius in the contracting phase would produce the same kind of power
spectrum as a quasi de Sitter Universe (an inflationary Universe)  when modes leave the Hubble radius in the expanding phase.
In fact, it has been shown that bouncing cosmologies in the flat FLRW space-time produce a nearly flat power spectrum \cite{spectrum}, as in inflation.

The main goal of the present work is to provide, for the ${\mathcal K}=1$-FLRW metric
and using a single scalar field,
background bouncing cosmologies in the framework of GR, and calculate the corresponding power spectrum of the curvature fluctuations in co-moving  coordinates.

These backgrounds cannot come from a field mimicking a fluid with Equation of State (EoS) $P=w\rho$ as in holonomy corrected LQC \cite{ewing}, because when the spatial curvature
is positive, a linear EoS produces cosmologies with a Big Bang and a Big Crush. Then,  the way to obtain bouncing cosmologies is to choose some particular bouncing backgrounds,
for instance $a(t)=(\rho_ct^2+1)^n$,
in our
case bouncing symmetric backgrounds that have a quasi-matter domination (see equation (\ref{f1}) which is our main model), that is,
we choose some Matter Bouncing Scenarios (see \cite{hc} for a recent review),  and apply the reconstruction
techniques to obtain a potential and the corresponding conservation equation (a second order differential equation) whose solutions lead to different cosmologies.
In general it is impossible to calculate analytically  that potential, and thus, numerical calculations are nedeed to recover  it. Once the potential has been  calculated,
one can calculate numerically
the different backgrounds, and for each one of them the corresponding  relevant terms of the power spectrum  such as the spectral index and its running, coming
from the Mukanov-Sasaki equation for geometries with positively curved space sections \cite{hn,Clesse,lilley}. However, these numerical calculations are very involved and
need future investigation, for this reason
here  we will only  calculate the spectral index and its running for our main background (\ref{f1}), and we will indicate how they will be for the other backgrounds.

\vspace{0.25cm}

The units used throughout the work are $\hbar=c=8\pi G=1$.


\section{Potential reconstruction for a single scalar field}
When one deals with a single scalar field, in the ${\mathcal K}=1$-FLRW geometry,  the Raychaudhuri equation in cosmic time becomes \cite{hn} (see also equation (6) of \cite{Clesse} where
it appears in conformal time)
\begin{eqnarray}\label{a10}
\dot{H}=-\frac{\dot{\varphi}^2}{2}+\frac{1}{a^2}.
\end{eqnarray}

Then, given some background, i.e. a scale factor $a(t)$, and thus the Hubble parameter $H(t)=\frac{\dot{a}(t)}{a(t)}$ and its derivative,  from the equation
\begin{eqnarray}\label{a11}
{\varphi}(t)=\int_{{\tilde t}_0}^t \sqrt{-2\left(\dot{H}(s)-\frac{1}{a^2(s)}  \right)}ds,
\end{eqnarray}
where ${\tilde t}_0$ is an arbitrary constant, we obtain the relation between the scalar field and the cosmic time, namely $\varphi=g(t)$.

On the other hand, using  the Friedmann equation \cite{fri},
\begin{eqnarray}\label{A11}
 H^2=\frac{\rho}{3}-\frac{1}{a^2},
\end{eqnarray}
we obtain the potential as a function of time
\begin{eqnarray}\label{a12}
\bar{V}(t)=3H^2(t)+\dot{H}(t)+\frac{2}{a^2(t)},
\end{eqnarray}
and making the replacement $t=g^{-1}(\varphi)$ (note that $g$ is always an inversible function because $\dot{g}\geq 0$ for all cosmic time $t$), one finally obtains
the corresponding potential
\begin{eqnarray}\label{a13}
V(\varphi)\equiv \bar{V}(g^{-1}(\varphi)).
\end{eqnarray}

In general, it is impossible to find analytically  the function $g^{-1}$, and thus, the potential must be obtained numerically, but there are some cases where an analytic calculation
is allowed.

\vspace{0.5cm}

{\bf Example $1$:}\label{ex:analitic}
 As an academic exemple we choose the scale factor $a(t)=a_0(\rho_ct^2+1)$ with $4a_0^2\rho_c=1$.
 One easily obtains $\dot{H}-\frac{1}{a}=-\frac{2\rho_c}{\rho_c t^2+1}$, then taking $\tilde{t_0}=0$ in (\ref{a11}) one gets
\begin{eqnarray}
 \varphi(t)=2\ln(\sqrt{\rho_c}t+\sqrt{\rho_c t^2+1})\Longleftrightarrow \rho_c t^2+1=\frac{(e^{\varphi}+1)^2}{4e^{\varphi}}.
\end{eqnarray}

\begin{figure}
  \includegraphics[width=0.50 \linewidth]{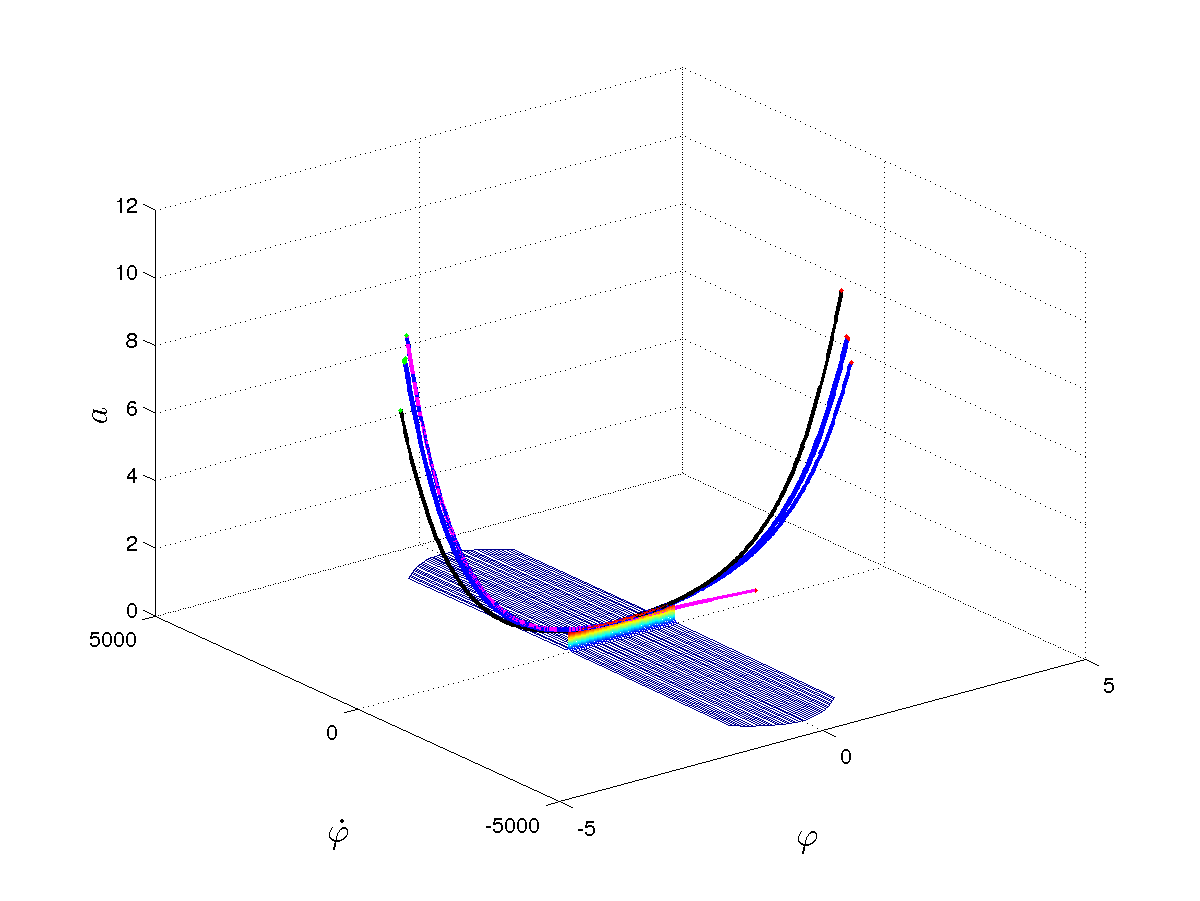}
 \includegraphics[width=0.50 \linewidth]{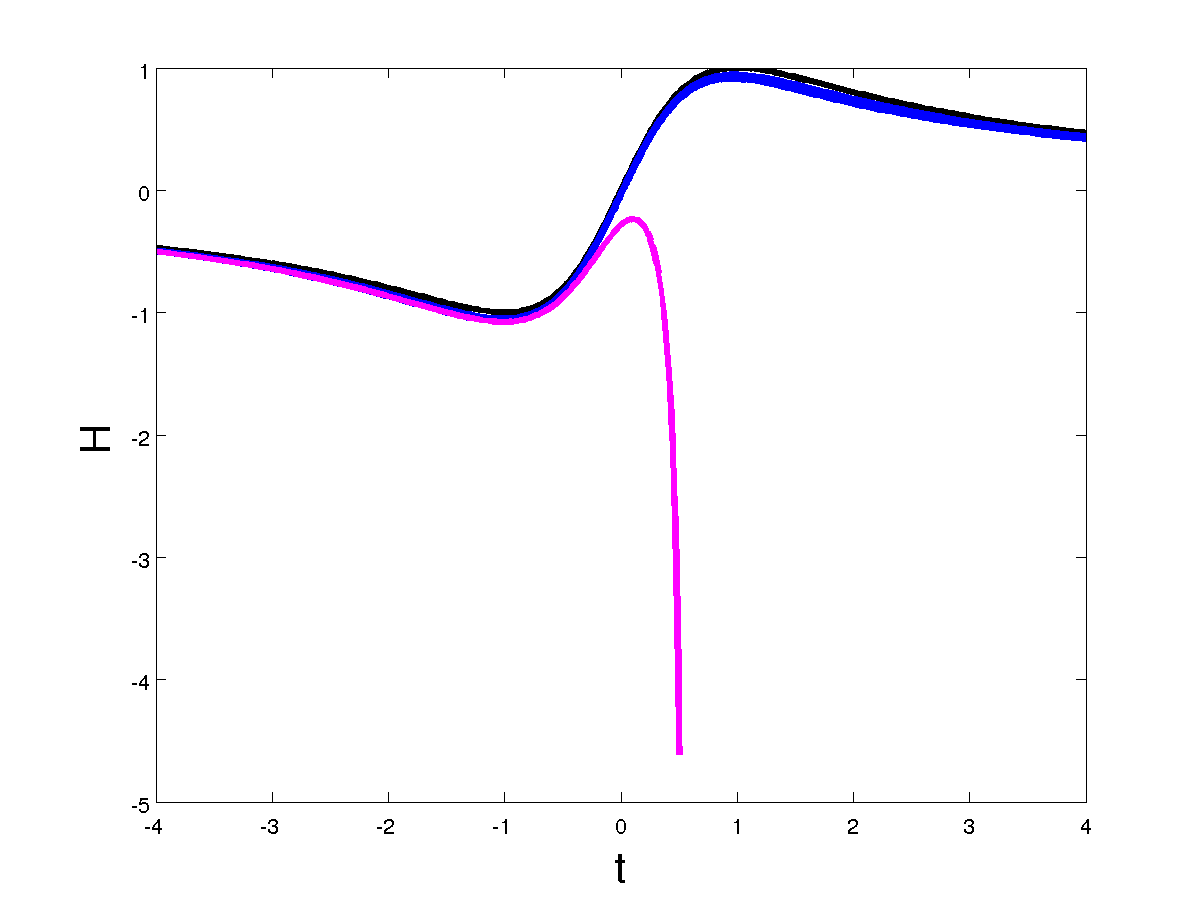}
\caption{The system presented in the Example $1$: its analytic solution (black), nearby solutions, which are all bouncing solutions in a neighbourhood of the
analytic solution (blue), and a nonbouncing solution (pink). (a): solutions $(\varphi(t),\dot{\varphi}(t),a(t))$ and bouncing surface. (b): $H(t)$ for each solution.}
\label{fig:exanalitic}
\end{figure}

Finally, inserting this last expression in (\ref{a12}) one will get
the symmetric potential
\begin{eqnarray}\label{p1}
 V(\varphi)=\frac{40\rho_c e^{\varphi}}{\left(1+e^{\varphi}\right)^2},
 \end{eqnarray}
{{}
which has the same shape as  the potential \cite{ewing} 
\begin{eqnarray}
V(\varphi)=\frac{2\rho_c(1-w) e^{\sqrt{3(1+w)}\varphi}}{\left(1+e^{\sqrt{3(1+w)}\varphi}\right)^2}
\end{eqnarray}
used in holonomy corrected 
LQC
to mimic a hydrodynamical fluid with EoS $P=w\rho$ (For the potential (\ref{p1}) one has to 
choose $w=-\frac{2}{3}$). Of course, in geometries with
positively  curved spatial sections
the potential (\ref{p1}) does not mimic  any hydrodynamical fluid with a linear EoS, but it depicts some bouncing backgrounds (see figure \ref{fig:exanalitic}).
}




\vspace{0.5cm}

Once we have reconstructed the potential, the dynamics is given by the following autonomous system
\begin{eqnarray}\label{a14}\left\{\begin{array}{ccc}
\dot{\varphi}&=&\psi\\
\dot{\psi}+3H_{\pm}(\varphi, \psi,a)\psi+V_{\varphi}&=&0\\
 \dot{a}&=&H_{\pm}(\varphi, \psi,a) a,\end{array}\right.
\end{eqnarray}
where
\begin{eqnarray}\label{a15}
H_{\pm}(\varphi, \psi,a)=\pm\sqrt{\frac{\psi^2}{6}+\frac{V(\varphi)}{3}-\frac{1}{a^2}}.
\end{eqnarray}

Note that equation (\ref{a14}) is a first order system of three differential equations, so apart from the originally chosen background it leads
to infinitely many new, different ones.

{{} In fact,
we have integrated numerically the equation (\ref{a14}) for the potential given in the Example $1$ (see figure \ref{fig:exanalitic}), obtaining a set of measure no zero in the ensemble  
of initial 
conditions $(\varphi_0,{\psi}_0,a_0)$ that leads to backgrounds with only one bounce, that is, depicting at very early times (resp. late times) a universe 
in the contracting (resp. expanding)
phase. As we have already explained this potential is a particular case of the potentials used in holonomy corrected LQC 
to mimic a hydrodynamical fluid with linear EoS. This opens the possibility to study these potentials in the context of ${\mathcal K}=1-$ FLRW geometry, and  to obtain 
new bouncing backgrounds solving numerically the equation (\ref{a14}).

Moreover, since the reconstruction is based in a given  bouncing  background, which is a solution of the system (\ref{a14}), it is clear that if one integrates
numerically (\ref{a14}) choosing initial conditions close to the
bouncing surface ${\frac{\psi^2}{6}+\frac{V(\varphi)}{3}-\frac{1}{a^2}}=0$ and to the chosen background, one will obtain some bouncing cosmologies. What one has to check numerically 
is their
viability in the sense that we will explain along the work.
}

\vspace{0.5cm}

{{} First of all, to obtain these viable bouncing backgrounds, we 
return to the reconstruction method, noting} that
the condition to reconstruct the potential is that $\dot{H}(t)-\frac{1}{a^2(t)}$ must be negative for all cosmic time. This places constrains on the backgrounds, for example
dealing with the simplest bouncing scale factor
$a(t)=a_0(\rho_ct^2+1)^{\frac{\alpha}{2}}$, only could be reconstructed when $\alpha a_0^2\rho_c\leq 1$. In fact, it is easy to show that the condition
\begin{eqnarray}\label{A1}
\alpha a_0^2\rho_c\leq 1, \quad  1\leq\alpha\leq 2,
\end{eqnarray}
is enough to reconstruct the potential corresponding to the simple background $a(t)=a_0(\rho_ct^2+1)^{\frac{\alpha}{2}}$.

On the other hand, for these backgrounds the effective EoS parameter given by the ratio of the pressure to the energy density is given by
\begin{eqnarray}\label{EoS}
 w\equiv \frac{P}{\rho}=-1-\frac{2(a^2\dot{H}-1)}{3(a^2H^2+1)}
 = -1-\frac{2}{3}\left(\frac{\alpha\rho_ca_0^2x^{\alpha}\left(\frac{2}{x^2}-\frac{1}{x}\right)-1}
 {\alpha^2\rho_ca_0^2x^{\alpha}\left(\frac{1}{x}-\frac{1}{x^2}\right)+1}\right),
\end{eqnarray}
where $x\equiv \rho_ct^2+1$.
When $x\gg 1$, i.e. far away from the bounce,  it becomes
\begin{eqnarray}
 w=-1+\frac{2}{3}\left(\frac{\alpha\rho_ca_0^2x^{\alpha-1}-1}
 {\alpha^2\rho_ca_0^2x^{\alpha-1}+1}\right).
\end{eqnarray}

Then, for $0<\alpha<1$, when $\alpha\rho_ca_0^2 \leq 1$ one has $w=-\frac{1}{3}$ and when $ \alpha\rho_ca_0^2\gg x^{1-\alpha}\gg 1$ one has
$w=-1+\frac{2}{3\alpha}$ which is nearly zero, and thus defines a quasi-matter dominated Universe, only when $\alpha\cong\frac{2}{3}$. On the other hand, for $\alpha\geq 1$
its impossible to have, far away to the bounce, a quasi-matter domination, because in that case one always has $w\leq -1+\frac{2}{3}=-\frac{1}{3}$.

This result means that one cannot reconstruct, using  a single scalar field, a bouncing cosmology with the simplest scale factor
$a(t)=a_0(\rho_ct^2+1)^{\frac{\alpha}{2}}$ ($\alpha>0$) that has a matter
domination period, because as we have already seen,  the reconstruction only holds for $\alpha\rho_ca_0^2\leq 1$ and matter domination requires, in that case,
$\alpha\rho_ca_0^2\gg 1$. For this reason, in the framework of GR, if one wants to reconstruct a bouncing cosmology containing a quasi
matter domination period in the contracting phase,
one has to improve the Matter Bounce Scenario. The most natural way is to introduce  a phase transition from the matter
domination to another regime.
Effectively, for the metric   $a(t)=\bar{a}_0(\bar{\rho}_ct^2+1)^{\frac{\bar{\alpha}}{2}}$ with
$1\leq\bar{\alpha}\leq 2$, it is possible to have, for values satisfying
$\bar{\alpha}\bar{\rho}_c\bar{a}_0^2\leq 1$, the conditions $|{a}H|\gg 1$ and $|a^2\dot{H}|\gg 1$ at early times. Then,  before this phase an scale factor of
the form $a(t)={a}_0\left(\frac{H_0}{\alpha}(t-t_0)+1  \right)^{\alpha}$ with $\alpha\cong \frac{2}{3}$
will satisfy $|{a}H|\gg 1$ and $|a^2\dot{H}|\gg 1$ and consequently, from (\ref{EoS}) one will obtain $w\cong 0$.

In fact, the following scale factor modelling a symmetric bounce with a phase transition
\begin{eqnarray}\label{f1}
a(t)=\left\{\begin{array}{ccc}
a_0\left(\frac{H_0}{\alpha}(t-t_0)+1    \right)^{\alpha}& \mbox{for}& t\leq t_0\\
\bar{a}_0(\bar{\rho}_ct^2+1)^{\frac{\bar{\alpha}}{2}}& \mbox{for}& t_0\leq t\leq |t_0|\\
a_0\left(\frac{|H_0|}{\alpha}(t-|t_0|)+1    \right)^{\alpha}& \mbox{for}&  |t_0|\leq t,
\end{array}\right.
\end{eqnarray}
with $t_0<0$, $\alpha\cong\frac{2}{3}$,
$a_0\equiv \bar{a}_0(\bar{\rho}_ct_0^2+1)^{\frac{\bar{\alpha}}{2}}$ and
$H_0\equiv \frac{\bar{\alpha}\bar{\rho}_ct_0}{\bar{\rho}_ct_0^2+1}$, satisfies all the conditions
if we assume
\begin{eqnarray}\label{f3}
1\leq\bar{\alpha}\leq 2,\quad \bar{\rho}_c \lesssim\frac{1}{\bar{\alpha}\bar{a}_0^2}\quad  \mbox{and} \quad|t_0|\gg \bar{a}_0.
\end{eqnarray}

The potential corresponding to the background (\ref{f1}) could not be analytically reconstructed and numeric calculations are needed. However, one can have an idea of its shape realizing
that
at very early times $\frac{1}{a^2}\gg |\dot{H}|$, then from (\ref{a11}) and (\ref{A11}) one obtains $V(\varphi)\propto |\varphi|^{\frac{2\alpha}{\alpha-1}}$, during matter domination
it is well-known that one has an exponential  potential  given by $V(\varphi)\propto e^{-\sqrt{3}|\varphi|}$ \cite{hc} and finally, during the bounce, if
$\bar{\alpha}=2$ and $4\bar{a}_0^2\bar{\rho}_c=1$,
 it will be exactly equal to (\ref{p1}).

\vspace{1cm}

{{}
The following  remark is in order:
A phase transition breaking the adiabaticity
is needed to explain the current temperature of the Universe via reheating,
due to the gravitational particle production at the phase transition time. A similar situation happens, in the framework of LQC,  with the so-called Matter-Ekpyrotic Bouncing 
Scenario \cite{caiewing},
where a phase transition from matter domination to an ekpyrotic regime produces a sufficient amount of particles to thermalize the Universe and match, in the expanding phase, with the hot
Friedmann Universe
\cite{Quintin}.
}


\section{Scalar Perturbations} Fist of all, based in observational evidences it is important to take into account that for spatially positive curved spaces, the
range of the co-moving wave-number is
$10^2\lesssim k \lesssim 10^8$ \cite{lilley}.

The Mukhanov-Sasaki (MS) equation (see \cite{sasaki} for the deduction of this equation in  flat backgrounds) for  positively curved spatial sections is \cite{hn,Clesse,lilley}:
\begin{eqnarray}\label{a23}
v_k''+(k^2-V)v_k=0,
\end{eqnarray}
where the derivatives are taken with respect de conformal time, $k^2=n(n+2)$ with $n\in {\mathcal N}$ and
\begin{eqnarray}\label{a24} V=\frac{z''_k}{z_k}+3(1-c_s^2),
z_k=a\frac{\dot{\varphi}}{H\chi_k},  \chi_k^2=1-3\frac{1-c_s^2}{k^2},
\end{eqnarray}
being
$c_s^2\equiv
\frac{\dot{P}}{\dot{\rho}}
=
-1-\frac{2\ddot{\varphi}}{3H\dot{\varphi}},$
the square of the velocity of sound.

If the background is given by (\ref{f1}), for $t<t_0$ the conformal time and the scale factor are given by
\begin{eqnarray}\label{a25}
\hspace{-0.35cm}\eta=\frac{\alpha}{(1-\alpha)a_0H_0}\left(\frac{H_0(t-t_0)}{\alpha}+1   \right)^{1-\alpha}\Longrightarrow
a  \propto \eta^{\frac{\alpha}{1-\alpha}}.
\end{eqnarray}

To calculate the power spectrum, we  write $z_k$ as follows:
\begin{eqnarray}\label{a26}
 z_k=\frac{a}{{\mathcal H}}\sqrt{3({\mathcal H}^2+1)}\sqrt{\frac{1+w}{1-3\frac{1-c_s^2}{k^2}}},
\end{eqnarray}
where ${\mathcal H}\equiv aH=\dot{a}$, and we evaluate $w$ and $c_s^2$, giving as a result

\begin{eqnarray}\label{a27}
 w= -1+\frac{2}{3\alpha}\frac{e^{2\mathcal N}+\alpha}{e^{2\mathcal N}+1},\quad
 c_s^2= -1+\frac{2}{3\alpha}\frac{e^{2\mathcal N}+\alpha^2}{e^{2\mathcal N}+\alpha},
\end{eqnarray}
where, in analogy of \cite{wilson1}, we have introduced the parameter ${\mathcal N}\equiv \ln(|{\mathcal H}|)$. Their derivatives with respect ${\mathcal N}$ are
respectively
\begin{eqnarray}\label{a28}
 w_{{\mathcal N}}= \frac{4}{3\alpha}\frac{e^{2\mathcal N}(1-\alpha)}{\left(e^{2\mathcal N}+1\right)^2},\quad
 c_{s;{\mathcal N}}^2\cong \frac{4}{3}\frac{e^{2\mathcal N}(2-\alpha)}{\left(e^{2\mathcal N}+\alpha\right)^2}.
\end{eqnarray}

Then, near the Hubble radius crossing $|{\mathcal H}|\sim k\gg 1$, the geometry is as the flat FLRW metric, and one has
\begin{eqnarray}\label{a29}
 w\cong c_s^2\cong -1+\frac{2}{3\alpha}, \quad  w_{\mathcal N}\cong \frac{4(1-\alpha)}{3\alpha}e^{-2{\mathcal N}}.
\end{eqnarray}

\begin{remark}
 When $\alpha=\frac{2}{3}$ one obtains a matter dominated Universe. This does not happen at very early times where ${\mathcal H}\sim 0$, because  in that case,
  ($0<\alpha<1$), one always has $w\cong-\frac{1}{3}$.
\end{remark}

Moreover, for modes near to leave the Hubble radius, i.e. when ${\mathcal H}\sim k\gg 1$, one has $z_k\cong -a \sqrt{3(1+w)}$,
and a simple calculation,
where one disregards $w_{\mathcal N}^2 $, because $w_{\mathcal N}^2\ll w_{\mathcal N}\cong \frac{4(1-\alpha)}{3\alpha}e^{-2{\mathcal N}}\ll 1  $
and uses that ${\mathcal N}'=\frac{{\mathcal H}'}{\mathcal H}=-\frac{1}{\eta}$, leads to
\begin{eqnarray}\label{a31}
\hspace{-0.15cm}\frac{z''_k}{z_k}\cong
\left(\frac{\alpha(2\alpha-1)}{(1-\alpha)^2}+\frac{3\alpha(1-3\alpha)}{4(1-\alpha)}w_{{\mathcal N}}+ 3\alpha w_{{\mathcal N}{\mathcal N}} \right)\frac{1}{\eta^2}.
\end{eqnarray}

Assuming that    $\alpha=\frac{2}{3}(1+\epsilon)$ where  $\epsilon$ is an small dimensionless parameter ($|\epsilon|\ll 1$),  and thus  $w\cong 0$, i.e.,
considering a Universe nearly matter dominated when the modes
leave the Hubble radius, one has
\begin{eqnarray}\label{a32}
 \frac{z''_k}{z_k}\cong 2(1 +9\epsilon-\frac{3}{4}w_{\mathcal N}+ w_{{\mathcal N}{\mathcal N}})\frac{1}{\eta^2}.
\end{eqnarray}

Then, for those modes
   the MS equation becomes
\begin{eqnarray}\label{a34}
v_k''+\left(\bar{k}^2-\frac{1}{\eta^2}\left(\nu^2-\frac{1}{4} \right)\right)v_k=0,
\end{eqnarray}
where $\bar{k}^2=n^2+2n-3$ and
$$\nu\cong \frac{3}{2}\left(1+4\epsilon-\frac{1}{3} w_{\mathcal N}+\frac{4}{9}w_{{\mathcal N}{\mathcal N}}\right).$$

And consequently, as in the flat case \cite{wilson1,eho}, our model leads to an spectral index  and its running given by
\begin{eqnarray}\label{a35}
n_s-1\equiv 3-2\nu=-12\epsilon+ w_{\mathcal N}-\frac{4}{3}w_{{\mathcal N}{\mathcal N}},\nonumber \\
\alpha_s\equiv \frac{d(n_s-1)}{d{\mathcal N}}=w_{{\mathcal N} {\mathcal N}}-\frac{4}{3}w_{{\mathcal N}{\mathcal N}{\mathcal N}}.
\end{eqnarray}

Note that $w_{\mathcal N}\cong (\frac{2}{3}-3\epsilon )e^{-2{\mathcal N}}\lesssim 10^{-4}$ and also $|w_{{\mathcal N}{\mathcal N}}|\lesssim 10^{-4}$ ,
then since $n_{s}-1=-0.0397\pm 0.0073$ \cite{Ade} and the term
$ w_{\mathcal N}-\frac{4}{3}w_{{\mathcal N}{\mathcal N}{\mathcal N}}$ does not contribute significatively to the spectral index, $n_s-1\cong -12\epsilon$, and in order to match with
observational date one has to  choose
$\epsilon\cong 0.0033\sim 10^{-3}$.

On the other hand, from (\ref{a35}) we can see that
the running is given by $\alpha_s\cong -5e^{-2{\mathcal N}}$, which is negative and small, because  $10^2\lesssim {\mathcal H}\lesssim 10^8$ means
$$-5 \times 10^{-16}\lesssim \alpha_s \lesssim -5\times 10^{-4},
$$
which
enters in
the marginalized 95.5\% Confidence Level (the distance of the theoretical value of the running to its average observational datal is less than $2\sigma$), because
Planck's 2013 results  \cite{Ade} give an spectral index with running $\alpha_s=-0.0134\pm 0.0090$.

To check the stability of this result, one has to consider other backgrounds
obtained as a solutions of the conservation equation provided by the reconstructed potential that corresponds to (\ref{f1}).
Then, backgrounds close to (\ref{f1}) during matter domination will satisfy that $w\cong 0$ and is nearly constant.
A calculation similar to that performed above shows
\begin{eqnarray}\label{a36}
 \frac{z''_k}{z_k} \cong {\mathcal H}'+{\mathcal H}^2 +w_{\mathcal N}\frac{2{\mathcal H}'{\mathcal H}^2+ {\mathcal H}''{\mathcal H}-({\mathcal H}')^2}{2{\mathcal H}^2}
 +w_{{\mathcal N}{\mathcal N}}
 \frac{({\mathcal H}')^2}{2{\mathcal H}^2}.
\end{eqnarray}

 Since  $w$ is nearly constant, we integrate the Raychaudhuri equation, in conformal time,  for modes near to cross  the Hubble radius (${\mathcal H}\sim k\gg 1$)
\begin{eqnarray}\label{a37}
 \frac{{\mathcal H}'}{{\mathcal H}^2+1}=-\frac{1}{2}(1+3w)\Longrightarrow\frac{{\mathcal H}'}{{\mathcal H}^2}=-\frac{1}{2}(1+3w),
\end{eqnarray}
obtaining
\begin{eqnarray}\label{a38}
 {\mathcal H}=\frac{2}{(1+3w)\eta}\cong \frac{2}{\eta}(1-3w).
\end{eqnarray}

For this background dynamics and those  modes
   the MS equation becomes
\begin{eqnarray}\label{a40}
v_k''+\left(\bar{k}^2-\frac{1}{\eta^2}\left(\nu^2-\frac{1}{4} \right)\right)v_k=0,
\end{eqnarray}
with $\nu\cong \frac{3}{2}\left(1-4w-\frac{1}{3} w_{\mathcal N}+\frac{4}{9} w_{{\mathcal N} {\mathcal N}} \right)$.

Then,  quasi matter domination when modes leave the Hubble radius leads to an spectral index  and its running given by
\begin{eqnarray}\label{a41}
n_s-1\cong 12 w,\quad \alpha_s=12w_{\mathcal N}+w_{{\mathcal N} {\mathcal N}}-\frac{4}{3}w_{{\mathcal N}{\mathcal N}{\mathcal N}},
\end{eqnarray}
where we have assumed $|w_{\mathcal N} |\ll |w|$ and  $|w_{{\mathcal N} {\mathcal N}} |\ll |w|$.

Unfortunately, in general, given a bouncing background that depicts a phase transition from the quasi-matter domination to another regime during contraction, these quantities,
must be calculated numerically because as we have already explained it is impossible to find analytically the corresponding potential. However,
to have an intuition about their numerical study, one can use formulas (\ref{a27}) and (\ref{a28}), which correspond to the background (\ref{f1}), with $\alpha=\frac{2}{3}(1+\epsilon)$.
In this case one has
\begin{eqnarray}\label{a42}
w\cong-\frac{1}{3(e^{2\mathcal N}+1)}-\epsilon\frac{e^{2\mathcal N}}{e^{2\mathcal N}+1},\quad
w_{\mathcal N}\cong \frac{2}{3}\frac{e^{2\mathcal N}}{(e^{2\mathcal N}+1)^2}.
\end{eqnarray}

Since $e^{-2\mathcal N}\lesssim 10^{-4}$ because ${\mathcal H}\gtrsim 10^{2}$, if we choose $\epsilon\cong 0.0033$ one will have $w\cong-\epsilon$ and
$w_{\mathcal N}\cong \frac{2}{3}e^{-2\mathcal N}$, leading to
the following values of the spectral parameters:
\begin{eqnarray}
 n_s-1\cong -12 \epsilon, \quad \alpha_s\cong 3e^{-2\mathcal N}.
\end{eqnarray}

Note that, for  these backgrounds that have a quasi matter domination when modes leave the Hubble radius and with $w$ nearly constant, the spectral index is the same as when $w$ is constant.
The
difference appears in the running, which now is small but positive although, from Planck's 2013 data \cite{Ade}, it also belongs in the marginalized 95.5\% Confidence Level
because $3e^{-2\mathcal N}\lesssim 10^{-3}$.

\vspace{1cm}

{{}
To end this section we will calculate the amplitude of the power spectrum for scalar perturbations and we also see that it survives through the bouncing transition.
First of all, recall that the range of the co-moving wave-numbers is $10^1\lesssim k\lesssim 10^8$. On the other hand,  the MS equation (\ref{a23}) contains the term
\begin{eqnarray}
3(1-c_s^2)=6+\frac{\ddot{H}a^2+2H}{H(\dot{H}a^2-1)},
\end{eqnarray}
where the denominator only vanishes at the bouncing time, because form the Raychaudhuri equation (\ref{a10}) one has $a^2\dot{H}-1=-\frac{a^2 \dot{\varphi}^2}{2}<0$. However,
for our particular background (\ref{f1}),
near
 the bouncing time  $\ddot{H}=-\frac{2\bar\alpha\bar{\rho}_c^2t}{(\bar{\rho}_ct^2+1)^3}(3-\bar{\rho}_ct^2)$, meaning that at $t=0$ one has
 \begin{eqnarray}
3(1-c_s^2)=6-\frac{2(3\bar\rho_c\bar{a}_0^2-1)}{\bar\alpha\bar\rho_c\bar{a}_0^2  -1}.
\end{eqnarray}

Then, since
in order to have a bounce, we will need the condition  $\bar\alpha\bar\rho_c\bar{a}_0^2< 1$,  if we choose our parameters in the way that they satisfy
$\bar\alpha\bar\rho_c\bar{a}_0^2\ll 1$, near the bounce,  we will obtain
$3(1-c_s^2)\cong 4\ll k^2.$
Note that this is a particular feature of our model, because in others models, see for instance the one studied by Martin and Peter in \cite{Clesse}, this quantity diverges at the bouncing
point.

Moreover, during matter domination one has $c_s^2\cong -1+\frac{2}{3\alpha}$ (see formula (\ref{a27})), then we can deduce that, for the modes we are dealing with,  the condition
\begin{eqnarray}
 |3(1-c_s^2)|\ll k^2,
\end{eqnarray}
is always satisfied. This is the key point, because for our particular model and for modes  with co-moving wave-numbers in the  range $10^1\lesssim k\lesssim 10^8$, 
the MS equation (\ref{a23}) becomes, as in the flat case,
\begin{eqnarray}\label{ms}
 v_k''+\left(k^2-\frac{z_k''}{z_k}\right)v_k=0.
\end{eqnarray}

Consequently,  to obtain the power spectrum, we can follow  the reasoning  of \cite{hc}, which goes as follows:

 In the quasi-matter domination phase,
 assuming the initial conditions of primordial perturbations to be vacuum fluctuations, one then obtains the solution to Eq. (\ref{a40})
\begin{eqnarray}\label{hankel}
 v_k(\eta)=\frac{\sqrt{-\pi\eta}}{2}e^{i(1+2\nu)\frac{\pi}{4}}H^{(1)}_{\nu}(-k\eta) ~.
\end{eqnarray}

 For modes well outside of the Hubble radius $k|\eta|\ll 1$,
 one can disregard the term $k^2$ and the general solution of (\ref{ms}) is
 \begin{eqnarray}\label{AAA}
  v_k(\eta)=C_1(k)z_k(\eta)+ C_2(k)z_k(\eta)\int_{\eta_0}^{\eta}\frac{1}{z_k^2(\bar{\eta})}d\bar\eta,
 \end{eqnarray}
where $\eta_0$ is some fixed time.


To calculate explicitly $v_k(\eta)$ in (\ref{AAA}) we will choose a pivot scale, namely $k_*$, and let $\eta_*$ be the time  at which the pivot scale crosses the Hubble radius
in the contracting phase. Then we could write the scale factor as follows
\begin{eqnarray}\label{A1}
 a(\eta) = a_*\left(\frac{\eta}{\eta_*}\right)^{\frac{1}{2}+\nu} \cong \frac{k_*}{|H_*|} \left(\frac{k_*|\eta|}{2}\right)^{\frac{1}{2}+\nu} ~,
\end{eqnarray}
where $a_*$ and $H_*$ are the values of the scale factor and Hubble parameter, respectively, at the crossing time. The approximation on the r.h.s.  comes from the fact that
in the quasi-matter dominated contracting phase, we have $aH\cong \frac{2}{\eta}$. Hence, since in the quasi-matter domination one has
$z_k(\eta)=\sqrt{3(1+w)}a(\eta)\cong \sqrt{3}a(\eta)$, choosing $\eta_0=\eta_*$, the leading term of (\ref{AAA}) can be written as follows,
\begin{eqnarray}\label{BBB}
 v_k(\eta) \cong {3\sqrt{3}}\frac{k_*}{|H_*|} |\eta_*|^{-\frac{1}{2}-\nu} {C}_2(k) \left(z_k(\eta)\int_{\eta_*}^{\eta} \frac{d\bar\eta}{z_k^2(\bar\eta)}\right) ~.
\end{eqnarray}

For that modes well outside of the Hubble radius, the solution (\ref{hankel}) should match with (\ref{BBB}). Using the small argument approximation in the Hankel function, we find
that the curvature fluctuation in co-moving coordinates is
\begin{eqnarray}\label{Zeta}
 \xi_k(\eta)\equiv\frac{v_k}{z_k(\eta)}\cong -\frac{i}{16}\left(\frac{6}{k}\right)^{\frac{3}{2}}e^{i(1+2\nu)\frac{\pi}{4}}
 \frac{k_*^3}{|H_*|} \int_{\eta_*}^{\eta} \frac{d\bar\eta}{z_k^2(\bar\eta)} \left(\frac{k}{k_*} \right)^{\frac{3}{2}-\nu}.
\end{eqnarray}

Note that, this formula is analytic in the bouncing phase because the function  $\frac{1}{z_k^2}(\eta)$ is analytic through the non-singular bounce. Then, in the wave-length  approximation,
$\left|\frac{z_k''}{z_k}\right|\gg k^2$, formula (\ref{Zeta}) is always valid provided that the background was  smooth enough.

Finally,
these modes will re-enter the Hubble radius at late times in the expanding phase, when the universe will be matter dominated. Then, the power spectrum is given by
\begin{eqnarray}\label{aa1}
 {\mathcal P}_{\xi}(k)\cong\frac{27}{64\pi^2} \frac{k_*^6}{H_*^2} \left(\int_{\eta_*}^{\tilde{\eta}_*} \frac{d\bar\eta}{z^2(\bar\eta)}\right)^2 \left(\frac{k}{k_*} \right)^{3-2\nu} ~,
\end{eqnarray}
where $\tilde{\eta}_*$ is the time when the pivot scale re-enter in the Hubble radius.

\vspace{0.5cm}

The following  remark is in order:
  Note that, for our background (\ref{f1}),  the derivative of the Hubble parameter is discontinuous at the phase transition time $\pm t_0$. Then, in order to apply formula (\ref{aa1})
  we would need that this phase transition to be   smoother, this can easily be achieved imposing that the phase transition is not instantaneous. On the other hand, it has been recently 
  proved, in the context
  of LQC (see \cite{caiewing} for details), that an abrupt phase transition 
  does not affect to the spectral index and its running, only  the amplitude of the power spectrum could change.

}

\vspace{0.25cm}

\section{Metodology for numeric calculations}

The numerical calculations associated to this problem are very involved and need future investigations. Here we only describe the way
to perform a numerical study:
 First of all, one has to reconstruct the potential for a background dynamics that depicts a given  Matter Bouncing Scenario, for instance, the one
given in (\ref{f1}), where the condition $(\ref{f3})$ must be satisfied  but with $|t_0|> 10^8 \bar{a}_0$, in order that modes with co-moving wave-number in the
range between $10^2$ and $10^8$ leave the Hubble radius before the ekpyrotic phase.

For that potential we have to solve the conservation equation for different initial conditions,  obtaining
different backgrounds. Once one has these bouncing cosmologies, one has to choose those
which  depict a quasi matter dominated Universe when modes  with co-moving wave-number in the range $ 10^2\lesssim k\lesssim 10^8$ leave the Hubble radius, and then calculate
$w$,  $w_{\mathcal N}$, $w_{{\mathcal N}{\mathcal N}}$ and $w_{{\mathcal N}{\mathcal N}{\mathcal N}}$ for their models.

Next, if we choose an initial instant, namely $t_i$, satisfying $t_i\lesssim t_0<-10^8\bar{a}_0$, and we evaluate our background depicting the Matter Bouncing Scenario
at that time, we will obtain the values
\begin{eqnarray}\label{e1}
a(t_i)={a}_i, \quad\varphi(t_i)=0, \quad\dot{\varphi}(t_i)=\sqrt{-2\left(\dot{H}_i-\frac{1}{a_i^2}\right)},
\end{eqnarray}
where $\dot{H}_i=\dot{H}(t_i)$ and $\varphi(t_i)$ is zero because one always could choose the value of ${\tilde t}_0$ in (\ref{a11})  equal to $t_i$.

Once that values are obtained,
 one has to calculate numerically some
other solutions of the conservation equation with initial conditions near to (\ref{e1}). With those solutions one calculates numerically $w(t)$, $\dot{w}(t)$,\dots ,$\dddot{w}(t)$ and
$\dot{a}(t)$, \dots ,$\ddddot{a}(t)$, and for values satisfying $10^2\lesssim |\dot{a}|\lesssim 10^8$ the parameter $w$ will be negative and near zero, which
provides  the theoretical value of the spectral index.
 In the same way, $\alpha_s$ is calculated from the formulas
$w_{\mathcal N}=\frac{\dot{a}\dot{w}}{\ddot{a}}$,
$w_{{\mathcal N}{\mathcal N}}=\frac{\dot{a}}{\ddot{a}}\frac{d}{dt}{\left(\frac{\dot{a}\dot{w}}{\ddot{a}}\right)}$
and $w_{{\mathcal N}{\mathcal N}{\mathcal N}}=\frac{\dot{a}}{\ddot{a}}\frac{d}{dt}\left(\frac{\dot{a}}{\ddot{a}}\frac{d}{dt}{\left(\frac{\dot{a}\dot{w}}{\ddot{a}}\right)}\right)$. 
Finally, one has to compare the theoretical results obtained numerically using this methodology  with observational data, what give us those backgrounds that could
realistically depict our Universe.

\section{Conclusions}
Matter Bounce Scenario in spacetimes with positively spatial curvature has been studied in this work, showing that it could be
a viable alternative to the inflationary paradigm. Without going beyond General Relativity and dealing with a single scalar field
(nonconvential matter fields are not  needed because a bounce is allowed in geometries with positive spatial curvature), bouncing background
with a quasi matter domination in the contracting phase could be built using the reconstruction method applied  to some chosen analytic scale factors as  (\ref{f1}).
Moreover, the chosen background leads to
 an spectrum of cosmological perturbations
that match with current observational data, specifically, the spectral index and its running. Therefore,
other backgrounds obtained numerically that are close to the chosen one, would have to give  similar results. Unfortunately, this only can be checked numerically, which is not
 a trivial task, this is
the reason
why, in this work, we have only showed it heuristically, and have proposed a method to perform the numerical calculations.

\vspace{0.5cm}
{\bf Acknowledgments.--} I would like to thank Professors  J. Amor\'os, Y.F. Cai, E. Elizalde and S.D. Odintsov  for their valuable comments. This investigation has been supported
in part
by MINECO (Spain), project MTM2014-52402-C3-1-P.

\end{document}